\documentclass[twocolumn,showpacs,aps,pre,nofootinbib]{revtex4}

\usepackage{graphicx}
\usepackage{amsfonts, amsmath, amstext, amssymb, amsfonts, amsxtra}
\usepackage{braket}
\usepackage{bbm}
\usepackage{bbold}
\DeclareMathOperator{\tr}{tr}

\newcommand{\im}{{\rm i}}

\newcommand{\Di}{\mathcal{D}}
 
\newcommand{\Lop}{\mathcal{L}} 
\newcommand{\rhop}{\hat{\rho}} 
\newcommand{\Hop}{\hat{H}} 
\newcommand{\Gop}{\hat{\Gamma}}     
\newcommand{\Pop}{\mathcal{P}}     
\newcommand{\Sop}{{\hat{S}}}   
\newcommand{\bop}{\hat{b}} 
\newcommand{\bdop}{\hat{b}^{\dagger}} 
\newcommand{\nop}{\hat{n}}

\begin{document}

\title{Period-doubling in period-$1$ steady states}
\author{Reuben~R.~W.~Wang$^{1}$\footnote{These authors contributed equally to the work.}, Bo Xing$^{1}$\footnotemark[\value{footnote}], Gabriel~G.~Carlo$^2$, Dario~Poletti$^{1}$}
\affiliation{$^{1}$Engineering Prouct Development Pillar, Singapore University of Technology and Design, 8 Somapah Road, 487372 Singapore \\
$^{2}$Departamento de F\'isica, CNEA, Libertador 8250, (C1429BNP) Buenos Aires, Argentina} 
%\author{Reuben~Wang$^{1}$\footnote{These authors contributed equally to the work.}, Bo Xing$^{1}$\footnotemark[\value{footnote}], Gabriel~G.~Carlo$^2$, Dario~Poletti$^{1,3}$}
%\affiliation{$^{1}$Engineering Prouct Development Pillar, Singapore University of Technology and Design, 8 Somapah Road, 487372 Singapore \\
%$^{2}$Departamento de F\'isica, CNEA, Libertador 8250, (C1429BNP) Buenos Aires, Argentina\\ 
%$^{3}$MajuLab, CNRS-UNS-NUS-NTU International Joint Research Unit, UMI 3654, Singapore} 
\date{\today}

% 03.75.Gg Entanglement and decoherence in Bose-Einstein condensates
% 03.65.Yz Decoherence; open systems; quantum statistical methods
% 05.45.Mt Quantum chaos; semiclassical methods
% 42.50.Dv Quantum state engineering and measurements

\begin{abstract}
Nonlinear classical dissipative systems present a rich phenomenology in their ``route to chaos'', including period-doubling, i.e. the system evolves with a period which is twice that of the driving. However, typically the attractor of a periodically driven quantum open system evolves with a period which exactly matches that of the driving. Here we analyze a manybody open quantum system whose classical correspondent presents period-doubling. We show that by analysing the spectrum of the periodic propagator and by studying the dynamical correlations, it is possible to show the occurrence of period-doubling in the quantum (period-$1$) steady state. We also discuss that such systems are natural candidates for clean Floquet time crystals.           
\end{abstract}

\maketitle

\section{Introduction}\label{sec:intro}

Classical driven and dissipative systems present a varied typology of dynamical behaviors. In these systems it is possible to observe very different types of attractors: fixed points, limit cycles and chaotic attractors. For quantum systems, if in some limit they can be reliably described by classical equations of motion, it is also possible to observe signatures of these behaviors (see for example \cite{Ctitanovic}).   

An important type of driven dissipative systems are those for which the driving is time periodic. The steady state of such systems, when unique, has a periodicity which is given exactly by the period of the driving, even if the classical corresponding system presents period-doubling or is chaotic \cite{HartmannHanggi2017}. Hence these systems deserve further investigations.   

An important insight into quantum systems is given by two-time correlations. For instance, current-current correlations on a thermal state can be used to infer its linear response transport properties. % \cite{LinearResponse}. 
For the case of quantum steady states, it was shown that the two-time correlations of a dissipative engineered quantum state can be significantly different from those of the target state \cite{SciollaKollath2015}.     

Here we show that by analysing the two-time correlations of periodic steady states, with a period exactly given by the driving period, it is possible to observe a period-doubling in the evolution of the correlation. This occurs when the corresponding classical system is in a parameter regime for which period-doubling occurs and when the effective Planck constant is small enough that the quantum dynamics mimics the classical dynamics for long enough times.   
The presence of an underlying period-doubling classical dynamics also naturally allows to interpret these systems as clean Floquet time crystal \cite{FTC1, FTC2, Ueda, Saro}.

The paper is divided as follows: in section \ref{sec:model} we introduce the model, then we describe its bifurcation map in section \ref{sec:bifurcation}, analyze the spectrum of the periodic propagator in section \ref{sec:spectrum}, and show the presence of period-doubling in the steady state in section \ref{sec:doubling}. In section \ref{sec:stability} we discuss that the system is a natural example of a clean Floquet time crystal and finally in section \ref{sec:conclusions} we draw our conclusions.

\section{Model, periodic steady state and meanfield equations}\label{sec:model}   

We consider a double well potential with $N$ atoms which is periodically driven and under the influence of dissipation. The system is described by a master equation whose time-dependent generator $\Lop_t$, of Lindblad form \cite{lind,gorini, alicki,book}, is composed of two parts
\begin{equation}
\label{eq:lind}
\dot{\rhop} = \Lop_t(\rhop) = -\im [\Hop(t),\rhop] + \mathcal{D}(\rhop). 
\end{equation}   
Note that we have set $\hbar=1$. The first part of Eq.(\ref{eq:lind}) describes the Hamiltonian evolution of the system's
density operator $\rhop$, due to the Hamiltonian $\Hop(t)$. We consider a double-well whose Hamiltonian is 
\begin{align}
\Hop(t)=&-J \left( \bdop_1 \bop_2 + \bdop_2 \bop_1 \right) + \frac{U}{2} \sum_{j=1,2} \nop_j\left(\nop_j-1\right)\nonumber \\
&+\varepsilon(t)\left(\nop_2 - \nop_1\right) \label{eq:Hamiltonian}
\end{align}
where $\bop^{}_j$ ($\bdop_j$) annihilates (creates) a boson at site $j$, while $\nop_j=\bdop_j\bop^{}_j$.      
The Hamiltonian parameters are $J$, the tunneling amplitude, $U$, the interaction strength, and
$\varepsilon(t)$, the modulation of the local potential. The modulation $\varepsilon(t)$ is chosen to be periodic of period $T=2\pi/\omega$, i.e. $\varepsilon(t)=\varepsilon(t+T)=\mu_0+\mu_1\sin(\omega t)$, where
$\mu_0$ and $\mu_1$ are, respectively, a static and a dynamic energy offset
between the two sites. This double-well Hamiltonian has been investigated in both theoretical \cite{coherent, Vardi, Witthaut, PolettiKollath2012} and experimental \cite{oberthaler, ober1} works. 

The second part of $\Lop_t$ in Eq.(\ref{eq:lind}) describes the dissipative evolution due to the dissipator
\begin{equation}
\label{dissipator}
\Di(\rhop) = \gamma \left(2\Gop\rhop \Gop^\dagger - \{\Gop^\dagger \Gop,\rhop\}\right)
\end{equation} 
where $\gamma$ is the dissipative rate while the jump operator is given by \cite{DiehlZoller2008, zoller, existence},
\begin{equation}
\Gop=(\bdop_1 + \bdop_2)(\bop_1-\bop_2). \label{eq:jump}
\end{equation}
This model has been investigated in \cite{HartmannHanggi2017, altri}

It was shown in detail in \cite{HartmannHanggi2017} that, given the periodicity of $\Lop_t$, it is possible to generate a Floquet map $\Pop_F=\Pop_{0,T}$ where $\Pop_{t_1,t_2}=\mathcal{T}e^{\int_{t_1}^{t_2}\Lop_t d\tau}$ and $\mathcal{T}$ is the time-ordering operator. The fix point of this map is the periodic steady state of the system $\rhop_{s}(mT)$ where $m$ is an integer number \cite{unique}. To compute $\rhop_{s}(t)$ at times $t\ne mT$, it is sufficient to evolve $\rhop(0)$ from time $0$ to $t$ using Eq.(\ref{eq:lind}).    

An important insight into the dynamics of this system is obtained, especially for large number of particles, by studying the corresponding classical meanfield equations of motion. To compute them, it is convenient to first rewrite Eq.(\ref{eq:lind}) in terms of the spin operators
$\Sop_x=\frac{1}{2N}\left(\bop^{\dagger}_1 \bop_2 + \bop^{\dagger}_2 \bop_1\right)$,
$\Sop_y= -\frac{\im}{2N}\left(\bop^{\dagger}_1 \bop_2 - \bop^{\dagger}_2 \bop_1\right)$
and $\Sop_z=\frac{1}{2N}\left(\nop_1 - \nop_2\right)$ and study their evolution in the
Heisenberg picture \cite{book}. The commutator between these operators is $\left[\Sop_x,\Sop_y\right]=\im\frac{\Sop_z}{N}$ and cyclic permutations. This implies that as $N\rightarrow\infty$, these spin operators commute, resulting in classical equations of motion (see \cite{HartmannHanggi2017} for more details).  

Since $\braket{\Sop^2}=\braket{\Sop_z^2}+\braket{\Sop_x^2}+\braket{\Sop_y^2}$ is a constant of motion (we have used the notation $\braket{\hat{O}} =\tr [\rhop \hat{O}]$ for the expectation value of the operator $\hat{O}$), it is possible to write the meanfield equations of motion of the system in terms of two angle variables $\theta$ and $\phi$ defined by $\left(\braket{S_x},\braket{S_y},\braket{S_z}\right) = \frac{1}{2} [\cos(\varphi)\sin(\vartheta),\sin(\varphi)\sin(\vartheta), \cos(\vartheta) ]$. We thus get the equations of motion \cite{HartmannHanggi2017}
\begin{align}
\dot{\vartheta} &= 2J\sin(\varphi) + 4\gamma N \cos(\varphi)\cos(\vartheta) \label{eq:thetadot} \\
\dot{\varphi} &= 2J\frac{\cos(\vartheta)}{\sin(\vartheta)}\cos(\varphi) - 2\varepsilon(t) +UN \cos(\vartheta) - 4\gamma N \frac{\sin(\varphi)}{\sin(\vartheta)}. \nonumber
\end{align}

\section{Bifurcation map}\label{sec:bifurcation}      

\begin{figure}
\includegraphics[width=\columnwidth]{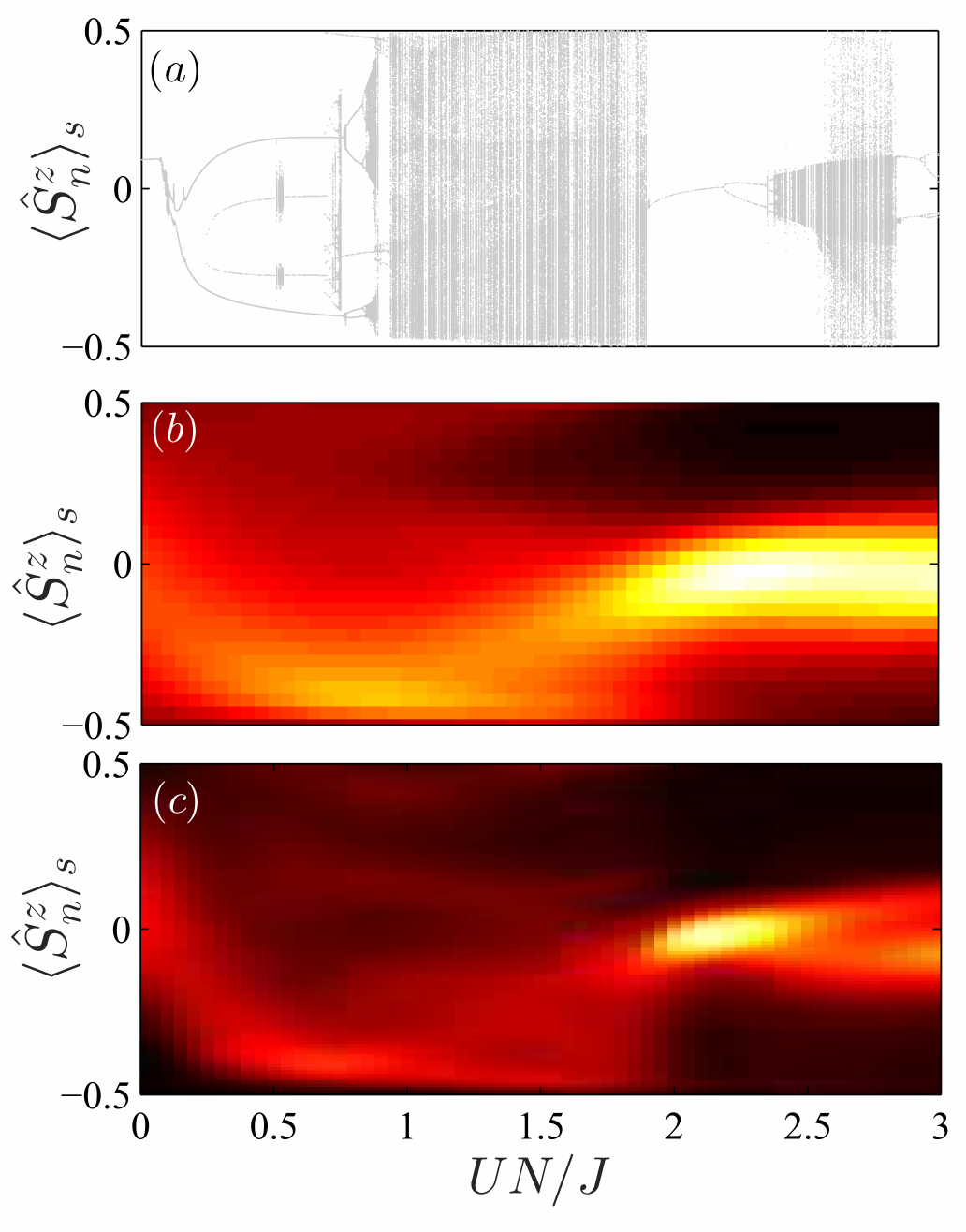}
\caption{(color online) Bifurcation maps for (a) classical meanfield equations (\ref{eq:thetadot}), (b-c) quantum system with repectively $N=25$ and $N=100$. The other parameters are $\mu_0=J$, $\mu_1=3.4J$ and $\gamma N=0.1 J$. } \label{fig:bifurcation}
\end{figure}

We analyze the quantum and classical bifurcation map for this system in Fig.\ref{fig:bifurcation} \cite{precedence}. More precisely, for the classical map we evolve Eq.(\ref{eq:thetadot}) for different initial conditions uniformly distributed over $\vartheta\in [-\pi,\pi]$ and $\phi\in [0,2\pi]$ for $t=800T$. We then record stroboscopically at times which are integer multiples of the driving period, $t=mT$, the value of $\vartheta$ (and hence of $\braket{\Sop^z}=1/2\cos(\vartheta)$) for the next $200$ (or more) periods and represent them in the Fig.\ref{fig:bifurcation}(a). 
For a similar interacting model \cite{Ivanchencko}, a quantum bifurcation map was produced using the trajectory method for the evolution of the density matrix \cite{MolmerCastin}. In Fig.\ref{fig:bifurcation}(b-d) we show the quantum bifurcation map from another approach. For any given value of the interaction we compute the steady state $\rhop_s(0)$ and then we project it over the eigenstates of $\Sop^z$ and take the trace. More precisely we can write $\Sop^z=\sum_n\Sop^z_n$ where $\Sop^z_n= (n/N - 1/2) | n,N-n\rangle\langle n, N-n|$ and the state $| n,N-n\rangle$ has $n$ particles in site $1$ and $N-n$ in site $2$. We thus compute $\braket{\Sop^z_n}_s=\tr\!\left( \Sop^z_n \rhop_s(0)\right)$ and produce an intensity plot as a function of both $\braket{\Sop^z_n}_s$ and the interaction strength $U/J$. We have used the notation $\langle\dots\rangle_s$ to remind the reader that the trace is taken over the steady state. In Fig.\ref{fig:bifurcation}(b,c) we show the bifurcation map respectively for $N=25$ and $N=100$ atoms. For the larger number of particles it is possible to see more clearly the underlying structure of the meanfield classical corresponding equations.

\section{Spectrum of the Floquet map}\label{sec:spectrum}  

\begin{figure}
\includegraphics[width=\columnwidth]{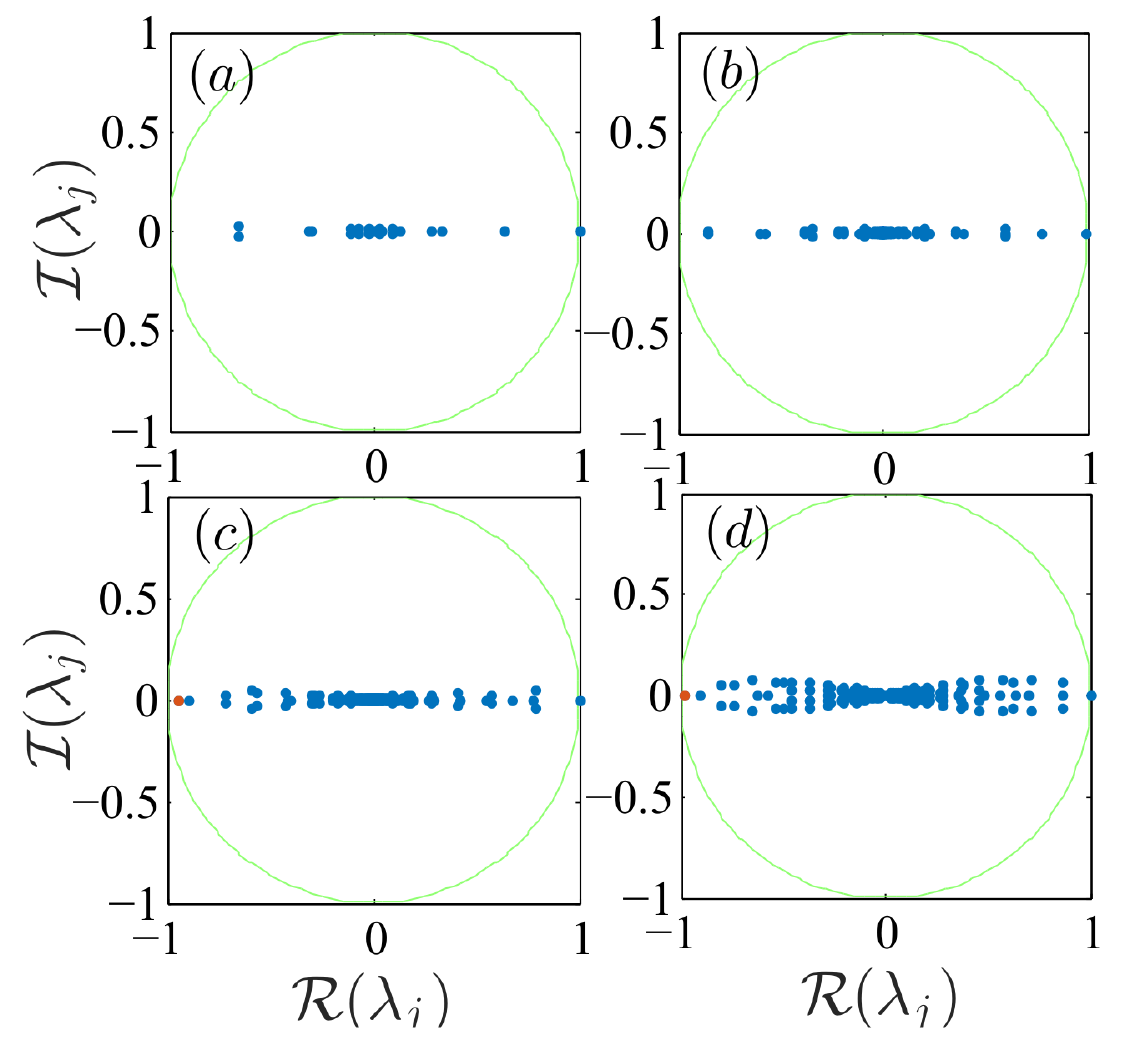}
\caption{(color online) Real and imaginary parts of the spectrum of the Floquet-rapidities for (a) $N=10$, (b) $N=25$, (c) $N=50$ and (d) $N=100$. The red dots in (c) and (d) represent the slowest decaying state. The other parameters are $\mu_0=J$, $\mu_1=3.4J$, $UN=0.2J$ and $\hbar\gamma N=0.1 J$.} \label{fig:spectrum}
\end{figure}

A signature of the presence of a bifurcation in the classical corresponding system leaves signatures in the spectrum of the Floquet map $\Pop_F$. We thus compute the eigenvalues of $\Pop_F$, which we refer to as `Floquet-rapidities' $\lambda_j$, for different particle numbers and we plot them in a complex plane in Fig.\ref{fig:spectrum}. In particular Figs.\ref{fig:spectrum}(a-d) correspond respectively to $N=10,\;25,\;50$ and $100$. Since the periodic steady state is unique, one Floquet-rapidity has exactly the value $1$. We note that as the number of particles increases, there is a Floquet-rapidity which approaches, but does not reach, the value $-1$ on the real axis (we highlight this point in red). The presence of such a slow decaying state with Floquet-rapidity $\approx -1$ indicates that dynamical properties can show an oscillatory behavior with a period which is twice that of the driving.

\section{Period-doubling}\label{sec:doubling}  

\begin{figure}
\includegraphics[width=\columnwidth]{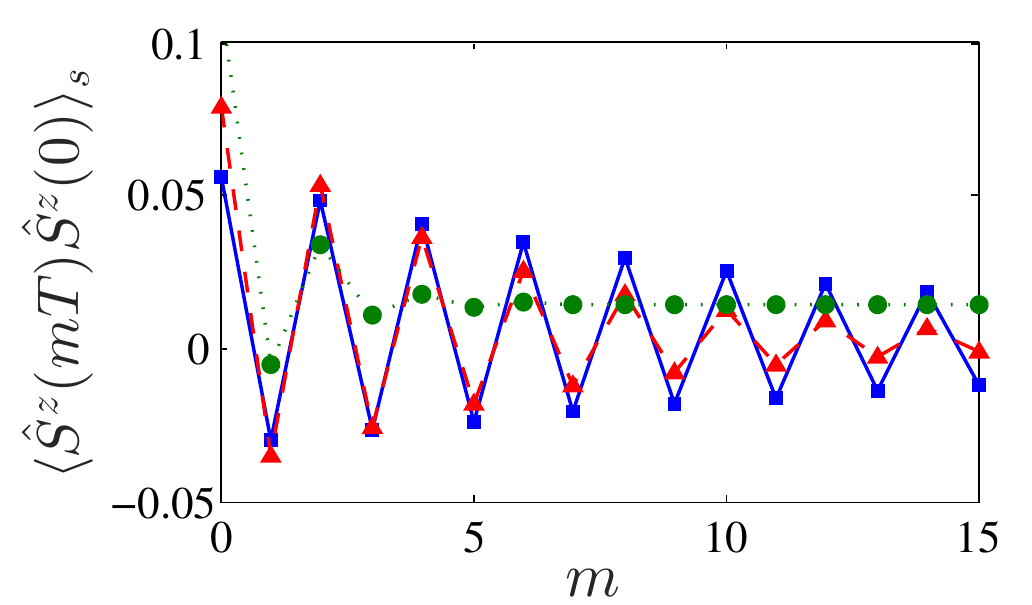}
\caption{(color online) Two-time correlation $\braket{\Sop^z(nT)\Sop^z(0)}_s$ versus time for a total particle number $N=5$, green circles, $N=25$, red triangles, and $N=100$, blue squares. Other parameters are $UN=0.2J$, $\mu_0=J$ $\mu_1=3.4J$ and $\gamma N=0.1J$.  } \label{fig:doubling}
\end{figure}

Despite the periodic steady state has a period which exactly matches that of the driving, if we observe dynamic correlations of the steady state we can find period-doubling. In particular we study the two-time correlation       
\begin{align}
\braket{\Sop^z(mT)\Sop^z(0)}_s=\tr(\Sop^z \left(\Pop_F\right)^m\Sop^z\rhop_s(0)),  
\end{align}
where, as a reminder, $\rhop_s(0)$ is the periodic steady state of of the Floquet map $\Pop_F$. We plot $\braket{\Sop^z(mT)\Sop^z(0)}_s$ in Fig.\ref{fig:doubling} as a function of time for different total particle number $N$. 
Fig.\ref{fig:doubling} demonstrates that the two-time correlation evolves with a period which is twice that of the driving, $T$, for an amount of time which becomes longer the more atoms are in the system. 
From our previous analysis of the Floquet-rapidities, we can understand this behavior from the presence of a Floquet-rapidity close to $-1$. In fact the long time evolution of the operator $\Sop^z\rhop_s(0)$, is dominated by the two slow decaying Floquet-rapidities at $1$ and close to $-1$. This last one induces the oscillation of the operator $\Sop^z\rhop_s(0)$ in time and results in the oscillation of the observable with period $2T$.        

\section{Bifurcation as driver of clean Floquet time crystals}\label{sec:stability}       

\begin{figure*}
\includegraphics[width=1.\textwidth]{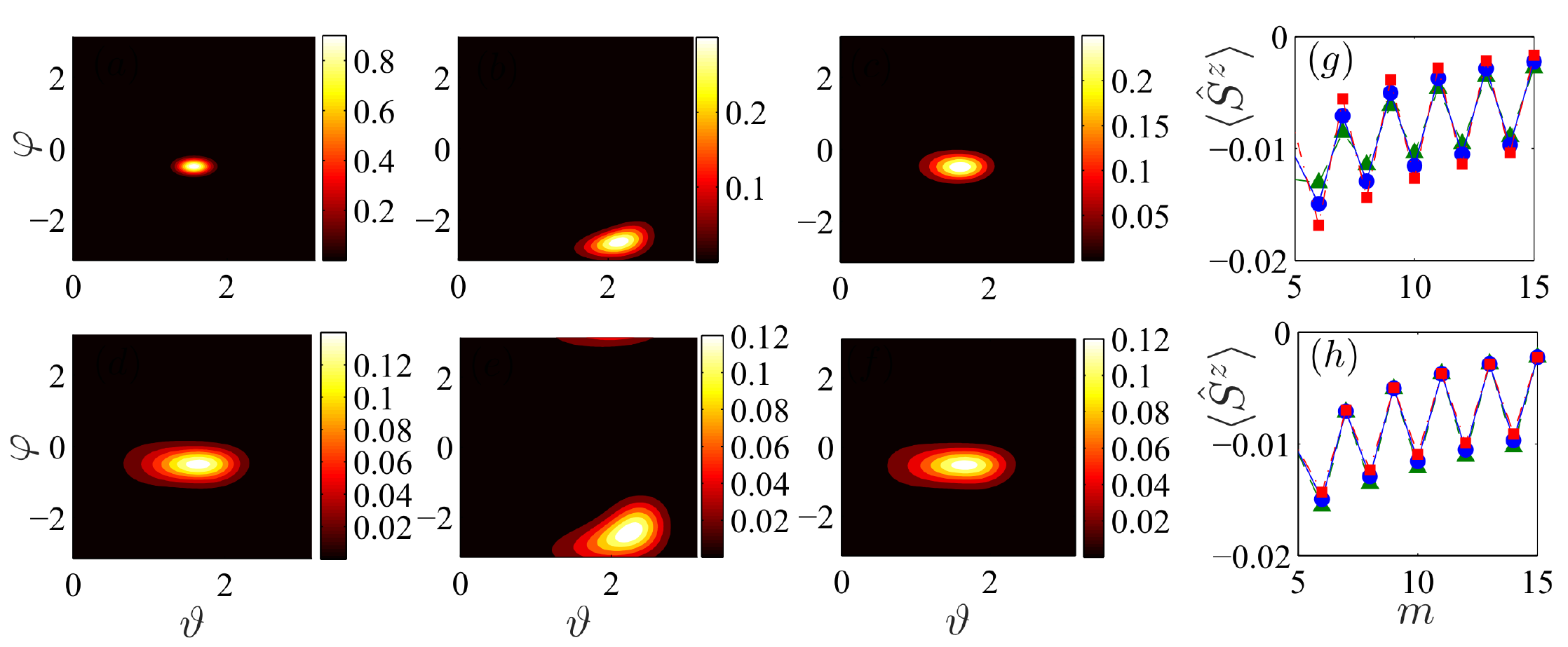}
\caption{(color online) (a-f) Husimi-Poincar\'e section of the evolution of a coherent state centered at one of the two classical periodic points of the classical Poincar\'e map. In particular the times are $t=0$ (a), $t=T$ (b), $t=2T$ (c), $t=6T$ (d), $t=7T$ (e), $t=8T$ (f). Panels (g-h) show $\langle \hat{S}^z\rangle$ at times given by different integers of period $t=mT$. In panel (g) we show that the evolution shows an alternating evolution for different initial conditions. In particular we chose coherent states centered in the points $(\vartheta,\varphi)=(2,-3)$ (blue circles), $(1.95,-3.05)$ (green triangles) and $(2.05,-2.95)$ (red squares). The strength of the interaction in (a-g) is $UN=0.2 J$. In panel (h) we show the robustness of the motion to different system parameters. In particular we evolve a coherent state centered at $(\vartheta,\varphi)=(2,-3)$ for parameters $UN=0.2 J$ (blue circles), $UN=0.21 J$ (green triangles) and $UN=0.19 J$ (red squares). Common parameters in (a-h) are $\mu_0=J$, $\mu_1=3.4J$, $\gamma N =0.1J$} \label{fig:stability}
\end{figure*}

Clean Floquet time crystals are defined as systems which in the thermodynamic limit fulfil the following properties: (i) there is a quantity which does not evolve with the period of the driving, (ii) it presents a periodic evolution without fine-tuning of the system parameters, and (iii) the periodic evolution should persist for an indefinitely long time \cite{FTC1}. If a quantum open periodically driven system has a classical correspondent which is in a period-doubling regime, then it is straight-forward to show that there are initial conditions which are robust both to changes in their exact position and of the system parameters and which show an evolution with a period different from the driving in the thermodynamic limit. 
To show this we take a coherent state centered in one of the two periodic points of the classical Poincar\'e section from Eq.(\ref{eq:thetadot}) and we evolve it in time. The coherent state is given by \cite{coherent1}   
\begin{align}
|\phi(\vartheta,\varphi)\rangle = \sum_{n=0}^N f_{n}(\vartheta,\varphi)|n,N-n\rangle \label{eq:coherent}              
\end{align}
where $f_n=\sqrt{\binom{N}{n}}\left[cos \left(\frac{\vartheta}{2}\right)\right]^n\left[\sin\left(\frac{\vartheta}{2}\right)e^{\im\phi}\right]^{N-n}$. 
In Fig.\ref{fig:stability}(a-f) we show stroboscopic images of the state as it evolves in time. Each panel is a Poincar\'e-Husimi section obtained by projecting the evolving state over coherent states. In particular we show the Poincar\'e-Husimi section of the state for times $t=0,\; T, \;2T, \;6T, \;7T$ and $8T$. We can observe that the coherent state jumps between two different positions, which is the expected behavior for classical period-doubling. Due to the finite number of particles, we also observe a broadening of the state.        
In Fig.\ref{fig:stability}(g-h) we focus on the stability of this dynamics. In Fig.\ref{fig:stability}(g) we show the evolution of $\langle \Sop^z\rangle$ versus time for different initial condition close to one of the classical periodic points. In Fig.\ref{fig:stability}(h) instead we vary the system parameters, still within the region of classical period-doubling, and we observe that the evolution is stable. As expected, the dynamics of this system fulfils the properties of a clean Floquet time crystal \cite{FTC1}.

\section{Conclusions}\label{sec:conclusions}  
In periodically driven open quantum system the density operator of the steady state typically evolves with a period which exactly matches that of the driving. This implies that static observables measured on this steady states can also oscillate at the same period, but not at multiples of it. 
Here we have presented clear signatures of period-doubling in such steady states when studying its dynamical correlations, and we have shown that the period doubling is due to the dynamics of the underlying classical correspondent system. The occurrence of period-doubling can be predicted by the presence of a Floquet-rapidity which approaches $-1$ and can induce oscillations of period $2T$ in dynamic observables.             
We have also shown that this open manybody quantum system can behave as a clean Floquet time crystal. 
It should here be pointed out the in \cite{WatanabeOshikawa2015} the authors stressed the importance to study two-time correlations in order to identify time crystals (although they were not considering periodically driven systems). In our work we have found that the study of two-time correlations can be used to identify the (period-$1$) steady state itself as a Floquet time crystal.   
Future works could study the complete route to chaos in periodically driven manybody open quantum systems.   
 
In the final stages of the preparation of this work, a partially related article was posted \cite{Ueda}, which however does not study dynamical correlations.

{\it Acknowledgments:} This work was supported by the Air Force Office of Scientific Research under Award No. FA2386-16-1-4041. D.P. acknowledges discussions with S. Denisov, R. Fazio, M. Hartmann, A. Rosch and A. Russomanno.


\begin{thebibliography}{1000}
\bibitem{existence} In general, the periodic driving also affects the coupling of the system to the environment. However here we consider the case in which this effect has been compensated to give some time-independent jump operators. We stress that our results are general and not dependent of the particular evolution of the quantum open system, as long as the classical limit gives a rich dynamics with period-doubling and route to chaos.
\bibitem{unique} In our case the steady state is unique.
\end{thebibliography}
\end{document}